\documentclass[10pt, conference, letterpaper]{IEEEtran}
\usepackage{booktabs}
\usepackage{cite}
\usepackage{amsmath,amssymb,amsfonts}
\usepackage{algorithmic}
\usepackage{graphicx}
\usepackage{textcomp}
\usepackage{multirow}
\usepackage{url}
\usepackage{xcolor}
\usepackage[linesnumbered,ruled]{algorithm2e}
\usepackage{textcase}  
\usepackage{microtype}
\IEEEoverridecommandlockouts
\begin{document}
\title{M$^3$LLM: Model Context Protocol-aided Mixture of Vision Experts For Multimodal LLMs in Networks}

\author{
\IEEEauthorblockN{Yongjie Zeng\IEEEauthorrefmark{1} and Hongyang Du\IEEEauthorrefmark{2}
\thanks{\IEEEauthorrefmark{1}Work done during internship at HKU.}
\thanks{\IEEEauthorrefmark{2}Corresponding author.}}
\IEEEauthorblockA{\IEEEauthorrefmark{1}China University of Mining \& Technology, Beijing, China}
\IEEEauthorblockA{\IEEEauthorrefmark{2}The University of Hong Kong, Hong Kong SAR, China}
\IEEEauthorblockA{Email: jiejiejie772@gmail.com, duhy@eee.hku.hk}
}

\maketitle

\begin{abstract}
Current Multimodal Large Language Models (MLLMs) rely on centralized architectures and often suffer from poor alignment between the input task and their fixed visual encoding modules, which limits performance on diverse and dynamic visual tasks. With the increasing deployment of resource-efficient models on edge devices in wireless networks, a new opportunity emerges to dynamically use distributed vision experts for improved MLLM inference quality. To enable this, we propose M$^3$LLM, where the Model Context Protocol (MCP) coordinates a mixture of vision experts to achieve distributed MLLMs. Specifically, MCP is an open protocol that structures the input task context into interpretable representations, enabling wireless network-aware coordination between the central model backbone and edge-hosted vision experts. Based on the MCP representation, M$^3$LLM formulates vision expert routing as a joint optimization problem that balances task-expert semantic compatibility and channel performance. To solve the resulting gradient conflicts, we develop a dual-stream Soft Actor-Critic (SAC) algorithm with decoupled reward signals and introduce an Adaptive Stability Enhancement Module (ASEM) based on hierarchical Bayesian modeling to ensure effective routing. Experiments show that M$^3$LLM improves task accuracy, reduces communication cost, and enhances expert routing adaptability under dynamic wireless network conditions.
\end{abstract}


\section{Introduction}
Large Language Models (LLMs) have achieved remarkable progress in natural language understanding, with Transformer-based architectures demonstrating capabilities approaching human intelligence~\cite{vaswani2017attention}. This success in linguistic domains has driven efforts to extend LLMs beyond text, which has led to the emergence of Multimodal LLMs (MLLMs). MLLMs such as Flamingo~\cite{alayrac2022flamingo}, BLIP-2~\cite{li2023blip2}, and LLaVA~\cite{liu2023visual} align visual encoders with pre-trained LLMs to enable joint vision-language understanding. More recent advances, including GPT-4V~\cite{openai2023gpt4v}, Gemini~\cite{team2023gemini}, and Qwen-VL, further expand the scope of MLLMs by enabling complex multimodal reasoning. These capabilities have made MLLMs indispensable for applications ranging from medical image analysis to autonomous navigation\cite{add15}, where joint reasoning over visual and textual information is essential.

Despite the progress of MLLMs, their design remains centered on single-node architectures, which demand large memory and GPU resources and rely on fixed visual encoding modules used in the training process that may not be able to adapt to diverse downstream tasks~\cite{radford2021learning}. Specifically, state-of-the-art MLLMs often require tens of gigabytes of memory and rely on high-throughput GPU computation, making them unsuitable for devices with limited processing and energy capacity~\cite{huang2019gpipe}. Furthermore, even when deployment is feasible, using a fixed set of visual experts often fails to support the heterogeneous nature of downstream tasks~\cite{radford2021learning}. For example, recognizing objects in street-view images depends on robustness to scale, occlusion, and lighting variations~\cite{geiger2012kitti}, whereas analyzing medical scans demands fine-grained texture sensitivity and anatomical prior knowledge~\cite{irvin2019chexpert}. A shared vision encoder backbone trained on general-purpose datasets cannot effectively meet these divergent demands, resulting in inefficient inference and degraded task performance~\cite{bommasani2021opportunities}.

To overcome the limitations of monolithic architectures, recent work has explored modular approaches that decompose MLLM's vision encoder into collections of specialized vision experts. This trend toward modularity leverages Mixture-of-Experts (MoE) principles to enable task-specific optimization~\cite{shazeer2017outrageously}. For instance, MoVA~\cite{zong2024mova} introduces semantic routing among visual experts, while MoE-LLaVA focuses on efficient expert allocation strategies~\cite{lin2024moe}. Other efforts include Uni-MoE~\cite{add1} that enables unified multimodal processing, and SPHINX~\cite{add2} that enhances multiscale visual understanding. Although these modular designs improve flexibility and task alignment, the trained model is still deployed as an integrated whole on a single device, assuming downstream tasks will remain within the coverage of the vision encoders used during the MLLM pre-training process.

Recent advances in model compression and wireless networking technologies present new opportunities to overcome these limitations. Quantization and pruning techniques now allow specialized vision models to run efficiently on edge devices with minimal accuracy loss~\cite{jacob2018quantization}. Simultaneously, 5G and emerging 6G~\cite{du2023semantic} networks provide ultra-low latency and dynamic resource allocation capabilities. This convergence has created a distributed ecosystem in which diverse edge devices such as smartphones, drones, and edge servers can each contribute specialized visual processing capabilities. By coordinating these distributed resources intelligently, it becomes feasible to move beyond centralized architectures toward distributed MLLMs to dynamically access and select visual experts across the network based on task requirements. Realizing this vision of distributed, adaptive MLLMs requires solving two fundamental challenges:

\begin{itemize}
\item \textit{Interoperable expert connectivity:} Defining unified interfaces for invoking heterogeneous vision experts across devices with different platforms and capabilities.
\item \textit{Network-aware expert coordination:} Designing scheduling mechanisms that jointly consider task semantic requirements and dynamic network conditions to guide expert selection and resource allocation.
\end{itemize}
Fortunately, the recently standardized Model Context Protocol (MCP) provides a foundation for the first challenge by enabling structured communication between Artificial Intelligence (AI) components~\cite{anthropic2024mcp}. MCP defines a lightweight schema for context exchange, commonly used in LLM systems to advertise capabilities, describe tasks, and return results through a unified interface. Building on this, we extend MCP to create learnable context encodings that capture not only task semantics but also device capabilities and real-time channel states. Leveraging MCP, we address the first challenge of how the MLLM backbone selects suitable vision experts on different devices. MCP transforms the pool of available experts into a coherent mixture of vision experts' paradigms for MLLMs, where each expert's functionality, input, and output representations are dynamically aligned with the central model.

\begin{figure}[t]
\centering
\includegraphics[width=0.5\textwidth]{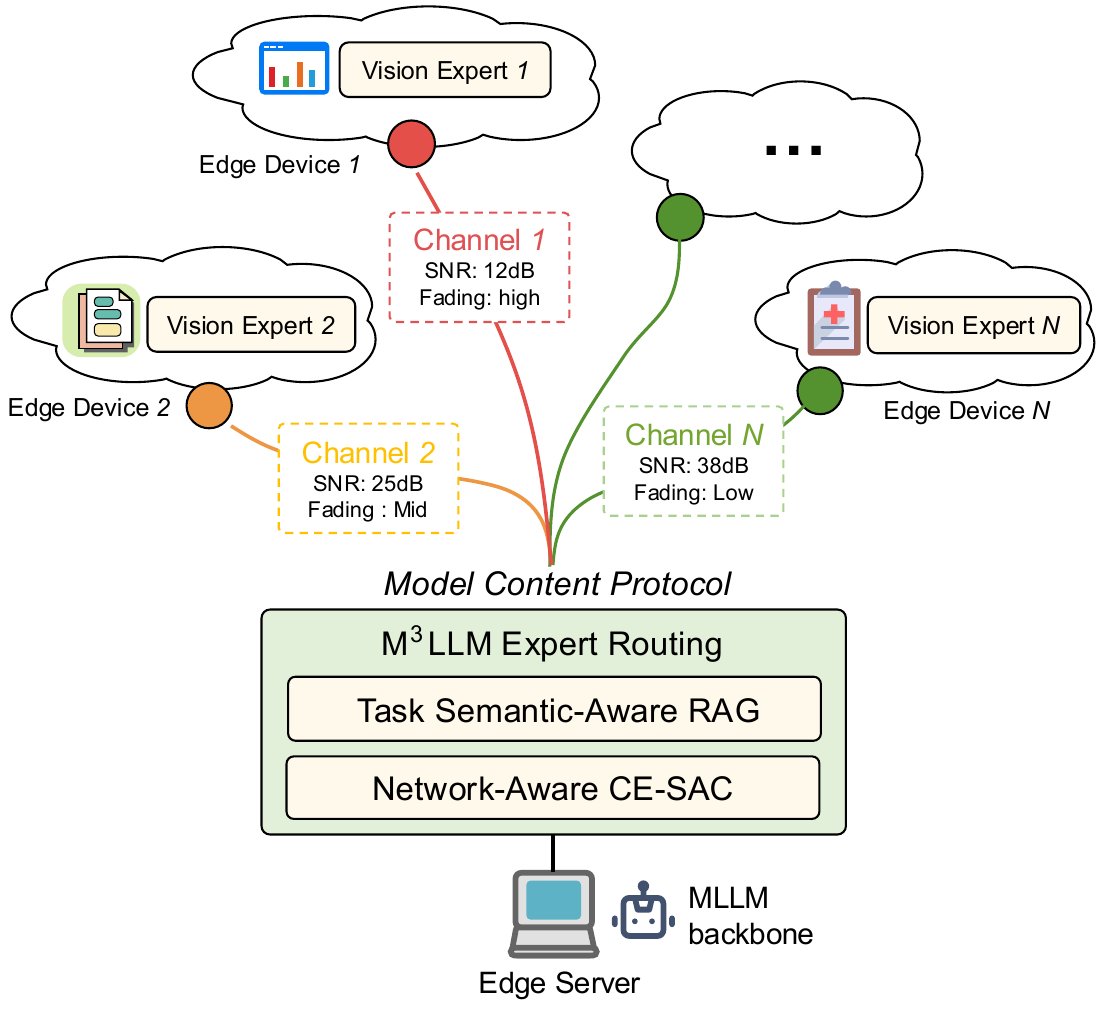}
\caption{The system model of M$^3$LLM in wireless networks.}
\label{fig:deployment_scenario}
\end{figure}

While the MCP paradigm facilitates semantic alignment between tasks and vision experts, selecting the most semantically compatible expert is not always optimal. In practice, the quality of the wireless link between the expert's host device and the central MLLM backbone significantly impacts the overall inference outcome. A highly relevant expert may produce precise embeddings, but if transmitted over a poor channel with high bit error rates, the resulting degradation can lead to worse performance than using a less optimal expert with a more reliable link. To address this second challenge, we further develop a dual-objective deep reinforcement learning (DRL) module based on Soft Actor-Critic (SAC)~\cite{haarnoja2018soft}, employing fully decoupled critics to eliminate gradient interference between semantic quality and communication robustness objectives. To better capture the impact of wireless dynamics on expert routing, we introduce an \textit{Adaptive Stability Enhancement Module (ASEM)} that models short-term channel volatility and long-term expert reliability through hierarchical Bayesian inference~\cite{blei2017variational}, providing stability priors that guide routing decisions under uncertainty. Our key contributions are summarized as follows:

\begin{itemize}
\item We propose M$^3$LLM, a distributed MLLM framework that enables the mixture of vision experts across multiple edge devices connected via wireless networks. This is achieved by extending the MCP to encode a unified context that integrates task semantics, device capabilities, and real-time channel statistics, supporting cross-device expert coordination.

\item We design Channel-Expert Soft Actor-Critic (CE-SAC), a dual-stream actor-critic architecture jointly optimizing expert selection and channel-aware routing. By leveraging separate critics for task-expert compatibility and wireless channel quality, CE-SAC eliminates reward interference and supports stable joint decision-making across semantic and communication dimensions.

\item We introduce ASEM, a hierarchical Bayesian state-space module capturing channel volatility and expert reliability distributions using variational inference. ASEM outputs latent stability priors that enhance SAC policy robustness against network noise and expert performance drift. Through extensive experiments, we show that M$^3$LLM achieves up to $51\%$ higher multimodal task accuracy under realistic wireless conditions compared to state-of-the-art MLLMs such as MoVA~\cite{zong2024mova}.
\end{itemize}

\section{System Architecture}
\label{sec:system_model}
In this section, we present the detailed architecture and methodology of M$^3$LLM. We first provide a high-level overview of the framework, followed by a detailed exposition of its two-stage hierarchical routing mechanism.

\subsection{Framework Overview}
M$^3$LLM is designed to coordinate specialized vision experts distributed across wireless edge networks. Each vision expert is hosted on a device with different capabilities and is trained for specific visual tasks. Instead of relying on a centralized backbone, M$^3$LLM dynamically selects and assembles experts based on arrived task semantics and runtime conditions, enabling flexible and efficient multimodal inference in heterogeneous distributed environments.

Central to our framework is MCP, which serves as the communication substrate enabling seamless interaction between distributed vision experts. It provides a standardized interface for capability advertisement, intermediate feature exchange, and response coordination across heterogeneous devices. To support scalable and context-aware expert selection, we integrate MCP with a lightweight Retrieval-Augmented Generation (RAG) mechanism. Specifically, RAG uses MCP-exposed capability descriptors to retrieve semantically relevant experts given an input query. This enables fast, interpretable filtering of the expert pool without invoking their underlying models. By coupling protocol-driven expert descriptions with retrieval-based preselection, M$^3$LLM achieves both modular extensibility and efficient routing in dynamic distributed environments.

\begin{figure*}[t]
\centering
\includegraphics[width=1.0\textwidth]{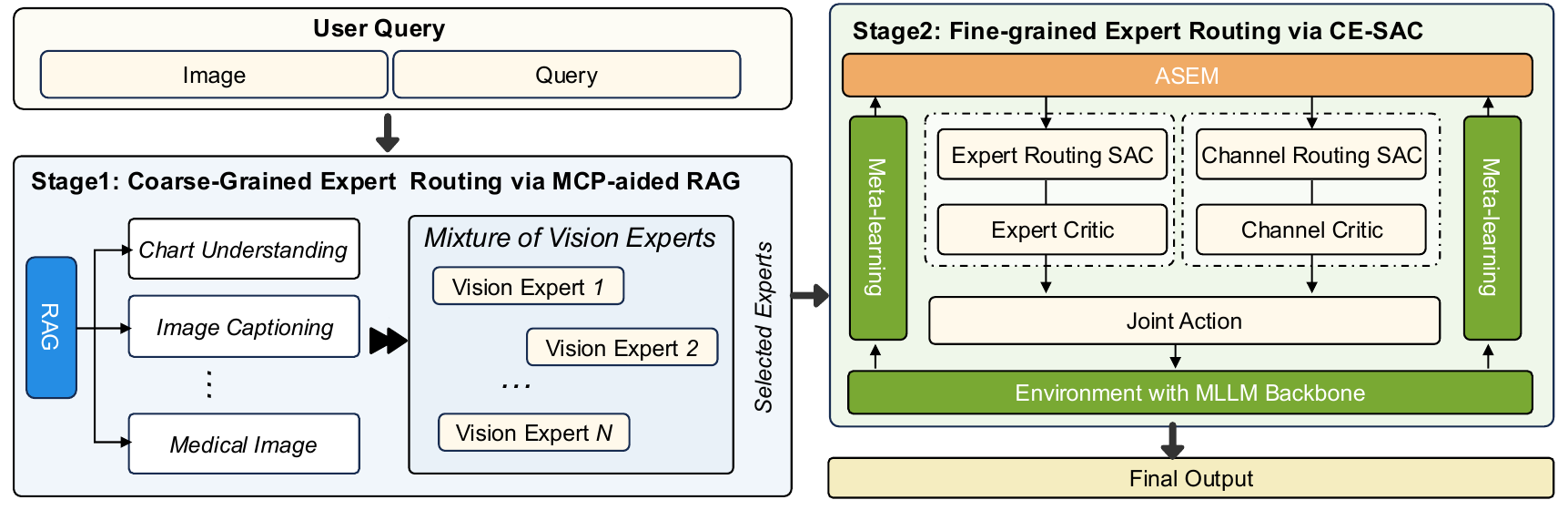}
\caption{Expert routing scheme of M$^3$LLM. Stage 1 performs coarse-grained expert filtering via MCP-aided RAG. Stage 2 executes network-aware fine-grained expert routing using a decoupled DRL agent, i.e., CE-SAC, which leverages a stability-aware state representation from ASEM.}
\label{fig:system_architecture}
\end{figure*}

As illustrated in Figure~\ref{fig:system_architecture}, the M$^3$LLM framework instantiates a distributed system where specialized vision experts collaborate through wireless networks to process multimodal queries. Each expert, which includes both general feature extractors and domain-specific analyzers, operates as an independent service accessible via MCP-compliant endpoints. When a user submits a multimodal query consisting of an image $\mathcal{I}$ and textual instruction $\mathcal{T}$, our M$^3$LLM orchestrates the optimal subset of experts to generate a comprehensive response. This orchestration must simultaneously optimize both task semantic alignment and communication robustness, necessitating a routing mechanism that can adapt to both query requirements and dynamic networks.
Therefore, we adopt a hierarchical design principle that decomposes the routing challenge into two complementary stages:
\begin{enumerate}
\item \textbf{Stage 1: Task Semantic-Aware Coarse-Grained Expert Routing via MCP-aided RAG.} This stage addresses a fundamental insight: not all vision experts are relevant to a given task. By enhancing RAG with MCP's standardized capability descriptors, we filter the expert pool based on semantic alignment. This ``semantic-first'' approach ensures that subsequent optimization focuses only on experts that can meaningfully contribute to the task, dramatically reducing the search space from potentially dozens of experts to a handful of relevant candidates.
\item \textbf{Stage 2: Network-Aware Fine-Grained Expert Routing via CE-SAC.} Semantic relevance alone is insufficient in distributed wireless environments. Experts are further selected while balancing inference quality against communication constraints, e.g., channel variability, latency, and resource limitations. A decoupled DRL architecture, i.e., CE-SAC, is used to prevent destructive gradient interference between conflicting objectives, enabling effective optimization under dynamic network conditions.
\end{enumerate}
By leveraging MCP-based communication and intelligent routing, our framework orchestrates specialized experts across the network edge to achieve superior flexibility, scalability, and performance compared to centralized models, while ensuring efficient resource utilization. The detailed procedures for both stages are presented in Section III.

\subsection{Wireless Network Environment}
We consider a practical wireless environment with $N$ vision experts, where each expert $e_i$ is accessible through a dedicated channel $c_i$ $(i = 1, \dots, N)$. This mapping reflects edge deployments in which experts reside on different physical nodes with independent network paths. Without loss of generality, we consider a parametric channel model that can later be specialized without altering our system design. Specifically, each channel is initialized with the following parameters:
\begin{align}
\mu_{{\rm SNR},i} &\sim \mathcal{N}(\mu_{\rm SNR}, \sigma_{\rm SNR}^2),\\
d_i &\sim \mathcal{N}(\mu_d, \sigma_d^2),\\
\sigma_{{\rm shadow},i} &\sim \mathcal{N}(\mu_{\rm shadow}, \sigma_{\rm shadow}^2),
\end{align}
where $\mathcal{N}(\mu,\sigma^2)$ denotes a Gaussian distribution with mean $\mu$ and variance $\sigma^2$. Here, $\mu_{{\rm SNR},i}$ is the mean Signal-to-Noise Ratio (SNR) of channel~$i$, $d_i$ is the physical distance between the edge device hosting expert~$i$ and the central coordinator hosting MLLM backbone, and $\sigma_{{\rm shadow},i}$ is the standard deviation of shadow fading for channel~$i$. These parameters establish a realistic wireless channel quality while showing significant variation across experts, reflecting differences in proximity, infrastructure, and network load.

Since wireless channels exhibit complex dynamics arising from path loss, shadowing, and multipath fading, we capture these effects through a hierarchical channel gain formulation that separates deterministic path loss from stochastic fading components~\cite{rappaport2002wireless}:
\begin{equation}
L_i(t) = L_0 + 10n\log_{10}\left(\frac{d_i}{d_0}\right) + X_{\sigma,i}(t),
\end{equation}
where $L_0$ represents the free-space path loss at reference distance $d_0$, $n$ is the path loss exponent characterizing the propagation environment, and $X_{\sigma,i}(t) \sim \mathcal{N}(0, \sigma_{{\rm shadow},i}^2)$ models the log-normal shadowing effect at time $t$. 
We consider the small-scale fading to capture multipath effects:
\begin{equation}
G_{{\rm SS},i}(t) = |h_i(t)|^2, \quad h_i(t) \sim \mathcal{CN}(\mu_h, \sigma_h^2),
\end{equation}
where $G_{{\rm SS},i}(t)$ denotes the small-scale fading power gain and $h_i(t)$ represents the complex channel coefficient following a circularly symmetric complex Gaussian distribution~\cite{goldsmith2005wireless}.
The composite channel gain determines the instantaneous SNR:
\begin{equation}
{\rm SNR}_i(t) = P_{\rm tx} - L_i(t) - 10\log_{10}\left( G_{{\rm SS},i}(t) \right) - N_0,
\end{equation}
where $P_{\rm tx}$ represents the transmission power, and $N_0$ is the thermal noise power spectral density~\cite{goldsmith2005wireless}.
We further consider the temporal correlation in shadowing through a first-order Gauss-Markov process~\cite{add3}:
\begin{equation}
X_{\sigma,i}(t+1) = \rho X_{\sigma,i}(t) + \sqrt{1-\rho^2}\sigma_{{\rm shadow},i}W_i(t),
\end{equation}
where $\rho$ represents the temporal correlation coefficient determining the memory of the shadowing process, and $W_i(t)$ is white Gaussian noise~\cite{stuber2017mobile}.

\subsection{M$^3$LLM Design Principles}\label{designp}
\subsubsection{Task Semantic-Aware Design Principle}
A central feature of M$^3$LLM is its ability to flexibly select vision experts distributed across multiple devices in wireless networks. Therefore, designing a task-driven evaluation framework for expert selection is essential. After each inference, we have the task type $\tau$ and the MLLM output $\mathcal{O}$ to construct a reward function that evaluates the quality of expert routing decisions, i.e., the expert weight distribution $\mathbf{w}_{\rm expert}$, as a task semantic-aware design principle:
\begin{equation}\label{tasksemreward}
R_{\rm LLM} = \sum_{i=1}^{4} \alpha_i \cdot R_i(\mathbf{w}_{\rm expert}, \tau, \mathcal{O}),
\end{equation}
where $\alpha_i$ are dimension weights reflecting the importance of each evaluation criterion. The components are defined as:
\begin{align}
R_1 &= \frac{1}{|\mathcal{E}_{\rm core}(\tau)|} \sum_{i \in \mathcal{E}_{\rm core}(\tau)} \mathbb{I}(\mathbf{w}_{\rm expert}[i] \geq \theta_{\rm act}),\\
R_2 &= 1 - \frac{1}{|\mathcal{E}_{\rm irr}(\tau)|} \sum_{i \in \mathcal{E}_{\rm irr}(\tau)} \mathbb{I}(\mathbf{w}_{\rm expert}[i] \geq \theta_{\rm sup}),\\
R_3 &= \frac{\sum_{i \in \mathcal{E}_{\rm core}(\tau)} \mathbf{w}_{\rm expert}[i]}{\sum_{i=1}^{N} \mathbf{w}_{\rm expert}[i]} \cdot \left(1 - H(\mathbf{w}_{\rm expert})\right),\\
R_4 &= \text{LLM}_{\rm eval}(\mathcal{O}, \tau, \{\mathbf{w}_{\rm expert}[i] : i \in \text{top-k}(\mathbf{w}_{\rm expert})\}),
\end{align}
where $\mathcal{E}_{\rm core}(\tau)$ and $\mathcal{E}_{\rm irr}(\tau)$ denote the task-critical and irrelevant expert sets, respectively, $\mathbb{I}(\cdot)$ is the indicator function, $\theta_{\rm act}$ and $\theta_{\rm sup}$ are activation and suppression thresholds, $H(\mathbf{w}_{\rm expert})$ is the entropy of the expert weight distribution, and $\text{LLM}_{\rm eval}$ invokes an external LLM to assess output consistency based on expert selection. Each component targets a specific aspect of semantic alignment. $R_1$ rewards correct activation of relevant experts, $R_2$ penalizes activation of irrelevant ones, $R_3$ encourages concentrated routing toward core experts with minimal dispersion, and $R_4$ ensures that the output aligns with the intended task semantics. 

\subsubsection{Wireless Network-Aware Design Principle}
While $R_{\rm LLM}$ captures task-expert semantic compatibility, it does not fully account for the impact of dynamic wireless environments on inference quality. To address this, we introduce a channel reward function that incorporates key network indicators, serving as the wireless network-aware design principle:
\begin{equation}\label{channelreward}
R_{\rm channel} = w_1 \bar{Q} + w_2 S + w_3 D + w_4 E,
\end{equation}
where
\begin{align}
\bar{Q} &= \frac{1}{|\mathcal{A}|}\sum_{i \in \mathcal{A}} \frac{{\rm SNR}_i(t) - {\rm SNR}_{\rm min}}{{\rm SNR}_{\rm max} - {\rm SNR}_{\rm min}},\\
S &= 1 - \frac{\sigma_{\rm SNR}(\mathcal{A})}{\bar{Q} + \epsilon},\\
D &= -\sum_{i \in \mathcal{A}} p_i \log p_i, \quad p_i = \frac{w_i}{\sum_j w_j},\\
E &= \frac{\sum_{i \in \mathcal{A}} \log_2(1 + {\rm SNR}_i(t))}{|\mathcal{A}| \cdot \log_2(1 + {\rm SNR}_{\rm max})},
\end{align}
$w_1$, $w_2$, $w_3$, and $w_4$ are weighting factors that balance the importance of different performance metrics, $\mathcal{A}$ denotes the set of active channels, $\bar{Q}$ represents the normalized average channel quality, $S$ captures channel stability through the coefficient of variation, $D$ measures load distribution entropy to encourage balanced utilization, $E$ indicates spectral efficiency based on Shannon capacity, ${\rm SNR}_{\rm min}$ and ${\rm SNR}_{\rm max}$ define the operational SNR range, $\sigma_{\rm SNR}(\mathcal{A})$ is the standard deviation of SNR values across active channels, $\epsilon$ is a small constant preventing division by zero, and $p_i$ represents the normalized weight assigned to channel $i$.
This reward structure ensures routing decisions optimize M$^3$LLM performance while considering long-term stability and system-wide quality.

\section{Methodology}
This section presents the MCP-aided expert retrieval pipeline, formalizes our two-stage methodology that comprises task-semantic-aware and network-aware expert routing, and proposes the stability-enhanced CE-SAC algorithm for M$^3$LLM.

\subsection{Stage 1: Semantic-Aware Coarse-Grained Expert Routing}
The first stage aims to efficiently prune the expert pool by identifying those semantically aligned with the user's multimodal query. As shown in Lines $1 - 9$ in Algorithm~\ref{algorithm1}, given an image $\mathcal{I}$ and a textual instruction $\mathcal{T}$, we employ a RAG pipeline that first uses the MLLM to extract a high-level task category tag from $\mathcal{T}$. This tag serves as a semantic query to retrieve relevant experts from a pre-built vector database, i.e., FAISS~\cite{add4}, which maps task categories to expert embeddings. The final output is a binary coarse mask:
\begin{equation}
\mathbf{m}_{\rm coarse} = \text{RAG}(\mathcal{I}, \mathcal{T}) \in \{0,1\}^N.
\end{equation}
where $\mathbf{m}_{\rm coarse}[i]=1$ indicates that expert $e_i$ is semantically relevant and should proceed to the next stage. This coarse filtering significantly reduces the candidate expert set, shrinking the search space from $2^N$ to a tractable subset and thereby reducing communication and computation costs in the fine-grained routing stage.

\begin{algorithm}[t]
\caption{Expert Routing in M$^3$LLM}
\label{algorithm1}
\let\oldnl\nl
\renewcommand{\nl}{}
\textbf{Input:} $\mathcal{I}$ (image), $\mathcal{T}$ (text instruction), $\mathcal{E} = \{e_1, \ldots, e_N\}$ (expert pool), $\mathcal{C}$ (channel states)\\
\textbf{Output:} $\hat{R}$ (optimal response), $\mathbf{w}^*$ (routing weights), $Q_{\text{total}}$ (quality score)
\let\nl\oldnl
\vspace{1mm}
Initialize $t = 1$, $\mathbf{w}^* = 0$, $Q_{\text{total}} = 0$\;
\While{$t = 1$}{
Encode context $\mathcal{M}_{ctx} \leftarrow \text{MCP}(\mathcal{I}, \mathcal{T})$\;
$\tau \leftarrow$ semantic tag extraction from $\mathcal{M}_{ctx}$\;
    
\If{semantic filtering required}{
$\mathbf{m}_{coarse} \leftarrow \text{RAG}(\mathcal{I}, \mathcal{T})$

\ForEach{$e_i \in \mathcal{E}$}
{
$\mathbf{m}[i] \leftarrow \mathbf{m}_{coarse}[i]$ 
}
        
$\mathcal{E}_{\text{cand}} \leftarrow \{e_i : \mathbf{m}[i] = 1\}$\;
}
Extract $\mathbf{f}_{img}, \mathbf{f}_{txt}$; compute $\mathbf{q}$ and $\mathbf{z}$\;
Form state $\mathbf{s} = [\mathbf{f}_{img}, \mathbf{f}_{txt}, \mathbf{m}, \mathbf{z}]$\;
$(\mathbf{w}_{\text{exp}}, \mathbf{w}_{\text{ch}}) \leftarrow \pi_{\text{CE-SAC}}(\mathbf{s})$\;
$\mathbf{w}^* \leftarrow \text{normalize}(\mathbf{w}_{\text{exp}} \odot \mathbf{w}_{\text{ch}} \odot \mathbf{m})$\;
\For{$i = 1$ to $N$}{
\If{$\mathbf{w}^*[i] > \epsilon$}{
Construct $\mathcal{M}_i$ and invoke $e_i$\;
$R_i \leftarrow e_i.\text{invoke}(\mathcal{M}_i, \text{channel}_i)$ (in parallel)\;
}
}
Aggregate response: $\hat{R} \leftarrow \sum \mathbf{w}^*[i] R_i$\;
Compute quality scores $Q_{\text{semantic}}, Q_{\text{channel}}$\;
$Q_{\text{total}} \leftarrow \alpha Q_{\text{semantic}} + \beta Q_{\text{channel}}$\;
$t \leftarrow t + 1$\;
}
\textbf{return} $\hat{R}, \mathbf{w}^*, Q_{\text{total}}$\;
\end{algorithm}

\subsection{Stage 2: Network-Aware Fine-Grained Expert Routing}
Following coarse-grained filtering, the fine-grained routing stage selects optimal experts from the candidate set by jointly considering semantic relevance and wireless channel conditions. We formulate this as a DRL problem and propose a custom agent, CE-SAC, with three key components: a stability-aware state representation module, i.e., ASEM, a decoupled actor-critic architecture, and an adaptive reward design.

\subsubsection{Stability-Aware State Representation with ASEM}
A major challenge in wireless environments is the high variability and noise in real-time channel observations~\cite{add7}, which can destabilize DRL policies. To address this, we introduce ASEM, a hierarchical Bayesian state-space model~\cite{add5} designed to extract stable latent representations from noisy inputs.
At each timestep $t$, ASEM processes four inputs: current channel observations $\mathbf{q}_t \in \mathbb{R}^N$, task semantic embedding $\boldsymbol{\tau}_t \in \mathbb{R}^{d_\tau}$ (from the user query), previous reward $r_{t-1} \in \mathbb{R}$, and the previous GRU hidden state $\mathbf{h}_{t-1} \in \mathbb{R}^{d_h}$. Using variational inference, ASEM infers two latent variables:
\begin{align}
\mathbf{z}_1^t &\in \mathbb{R}^{d_{z_1}} \text{ (short-term dynamics)},\\
\mathbf{z}_2^t &\in \mathbb{R}^{d_{z_2}} \text{ (long-term trends)},
\end{align}
where $d_{z_1}$ and $d_{z_2}$ are the dimensions of the short-term and long-term latent spaces, respectively. Short-term dynamics $\mathbf{z}_1^t$ capture immediate channel fluctuations and transient network events, while long-term trends $\mathbf{z}_2^t$ model persistent patterns in expert reliability and channel statistics. This hierarchical representation enables the system to respond to significant network changes while filtering out momentary disturbances.
Latent variables are inferred through variational approximation:
\begin{equation}
\mathcal{L}_{\rm ELBO} = \mathbb{E}_{q_\phi}[\log p_\theta(\mathbf{q}_t|\mathbf{z}_1^t, \mathbf{z}_2^t)] - D_{\rm KL}[q_\phi(\mathbf{z}^t|\mathbf{x}^t)||p(\mathbf{z}^t)],
\end{equation}
where $\mathcal{L}_{\rm ELBO}$ is the evidence lower bound, $q_\phi$ denotes the variational posterior parameterized by $\phi$, $p_\theta$ represents the generative model parameterized by $\theta$, and $D_{\rm KL}$ is the Kullback-Leibler divergence~\cite{add6}.
The resulting latent encoding $\mathbf{z}_{\rm ASEM} = [\mathbf{z}_1^t, \mathbf{z}_2^t]$ is concatenated with base features to form the comprehensive state representation: As shown in Lines 10-11 in Algorithm 1,
\begin{equation}
\mathbf{s}_t = [\mathbf{f}_{\rm img}^{d_{\rm img}}, \mathbf{f}_{\rm text}^{d_{\rm text}}, \mathbf{r}_{\rm tag}^{d_{\rm tag}}, \mathbf{m}_{\rm coarse}^{N}, \mathbf{z}_{\rm ASEM}^{d_{z_1}+d_{z_2}}] \in \mathbb{R}^{d_s},
\end{equation}
where $\mathbf{f}_{\rm img}$ denotes the image feature vector of dimension $d_{\rm img}$, $\mathbf{f}_{\rm text}$ represents the text feature vector of dimension $d_{\rm text}$, $\mathbf{r}_{\rm tag}$ is the task category representation of dimension $d_{\rm tag}$, and $d_s$ is the total state dimension.

By filtering transient noise and producing a stable, informative state representation, ASEM enables the DRL agent to learn reliable routing policies under dynamic and unpredictable wireless conditions.

\subsubsection{Channel-Expert Actor-Critic Architecture (CE-SAC)}
At the core of our DRL agent is CE-SAC, a dual-stream SAC architecture designed to resolve the conflict between optimizing semantic quality and wireless communication robustness. In conventional single-design principle settings, gradients from competing goals can interfere, destabilizing policy learning. To address this, CE-SAC adopts a decoupled design where the actor network extracts a shared state representation and outputs expert and channel weights through separate heads: As shown in Lines 12-21 in Algorithm 1,
\begin{equation}
\pi_\psi: \mathbf{s}_t \mapsto (\mathbf{w}_{\rm expert}, \mathbf{w}_{\rm channel}) \in [0,1]^N \times [0,1]^N,
\end{equation}
where $\pi_\psi$ denotes the policy network parameterized by $\psi$, $\mathbf{w}_{\rm expert}$ represents the expert selection weights, and $\mathbf{w}_{\rm channel}$ indicates the channel preference weights. Both heads utilize Gumbel-Softmax activation with temperature $\tau$, enabling differentiable sampling while maintaining exploration. The shared backbone ensures computational efficiency while the separate heads preserve objective independence.

To complement the actor, we design a dedicated quadruple-critic structure. Two expert critics, i.e., $Q_{\theta_1}^{\rm expert}$ and $Q_{\theta_2}^{\rm expert}$, are trained on task semantic quality rewards, i.e., \eqref{tasksemreward}:
\begin{equation}
\mathcal{L}_{e}(\theta_i) = \mathbb{E}_{(\mathbf{s},\mathbf{a},R_{\rm LLM},\mathbf{s}') \sim \mathcal{D}}\left[\frac{1}{2}\left(Q_{\theta_i}^{e}(\mathbf{s},\mathbf{a}) - y_{e}\right)^2\right],
\end{equation}
where $y_{e} = R_{\rm LLM} + \gamma \min_j Q_{\theta_j}^{e}(\mathbf{s}', \mathbf{a}')$, $\mathcal{D}$ denotes the replay buffer, $\mathbf{a}$ represents the action taken, $R_{\rm LLM}$ is the semantic quality reward, $\mathbf{s}'$ is the next state, and $\gamma$ is the discount factor.

Similarly, two channel critics, i.e., $(Q_{\phi_1}^{\rm channel}$ and $Q_{\phi_2}^{\rm channel})$, are trained using communication quality reward, i.e., \eqref{channelreward}:
\begin{equation}
\mathcal{L}_{c}(\phi_i) = \mathbb{E}_{(\mathbf{s},\mathbf{a},R_{channel},\mathbf{s}') \sim \mathcal{D}}\left[\frac{1}{2}\left(Q_{\phi_i}^{c}(\mathbf{s},\mathbf{a}) - y_{c}\right)^2\right],
\end{equation}
where $R_{\rm channel}$ denotes the communication quality reward.

This separation prevents gradient interference between two design principles as we discussed in Section~\ref{designp}, enabling stable convergence and balanced policy learning without manual weighting.
Final routing weights combine expert and channel decisions:
\begin{equation}
\mathbf{w}_{\rm final} = \frac{\mathbf{w}_{\rm expert} \odot \mathbf{w}_{\rm channel}}{||\mathbf{w}_{\rm expert} \odot \mathbf{w}_{\rm channel}||_1 + \epsilon},
\end{equation}
where $\odot$ denotes element-wise multiplication and $\epsilon$ is a small constant for numerical stability.

This multiplicative formulation enforces agreement between task-expert compatibility and wireless channel quality, ensuring that only experts satisfying both criteria are prioritized. Moreover, the $\ell_1$ normalization produces a valid probability distribution, enabling coherent response aggregation and improving interpretability during inference.

\subsubsection{Reward Decomposition and Meta-Analysis}
The reward design maintains separation between task semantic-aware and wireless network-aware design principles while incorporating meta-level performance analysis. Task semantic rewards $R_{\rm LLM}$ evaluate response accuracy, task relevance, and output coherence using LLM feedback. Communication rewards $R_{\rm channel}$ capture network-related metrics such as SNR, channel stability, and resource usage.

Although the expert and channel critics are trained independently on their respective rewards, we compute a combined score $R_{\rm final} = \alpha R_{\rm LLM} + \beta R_{\rm channel}$ as system overall performance. Here, $\alpha$ and $\beta$ denote the relative weights reflecting the practical trade-off between inference quality and communication reliability. This combined metric does not influence policy learning.

To assess the reliability of each reward signal under varying conditions, we also introduce a lightweight meta-analysis module. It processes confidence-related features from both reward streams and extracts a $d_{\rm conf}$-dimensional vector $\mathbf{f}_{\rm conf}$ representing reward statistics, temporal consistency, and environmental dynamics. A small neural network maps $\mathbf{f}_{\rm conf}$ to reliability weights $[w_{\rm llm}, w_{\rm channel}]$, which indicate the trustworthiness of the respective signals. While excluded from training updates, these weights support system diagnostics and performance monitoring.

\begin{figure*}[t]
\centering
\includegraphics[width=\textwidth]{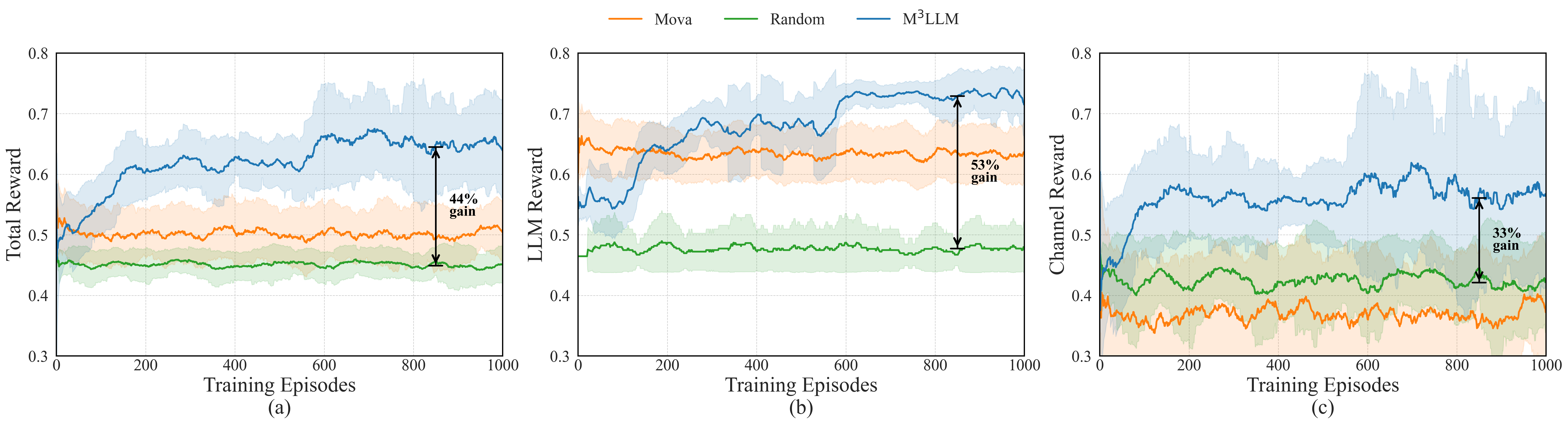}
\caption{Training dynamics comparison across $1,000$ episodes. (a) Total reward evolution showing M$^3$LLM's superior convergence and final performance. (b) LLM reward demonstrating consistent semantic quality improvements. (c) Channel reward highlighting M$^3$LLM's unique ability to optimize network quality while baselines remain static. Shaded areas represent confidence intervals.}
\label{fig:training_dynamics}
\end{figure*}
\begin{table*}[t]
	\centering
	\caption{Comprehensive performance comparison on the MME benchmark.}
	\label{tab:main_results}
	\begin{tabular}{lcccccc}
		\toprule
		\multirow{3}{*}{\textbf{Method}} & \multicolumn{3}{c}{\textbf{Task Semantic-aware Performance}} & \multicolumn{3}{c}{\textbf{Wireless Network-aware Performance}} \\
		\cmidrule(lr){2-4} \cmidrule(lr){5-7}
		& \textbf{LLM Quality} & \textbf{Task-Expert} & \textbf{Expert} & \textbf{Channel} & \textbf{SNR} & \textbf{Channel} \\
		& \textbf{Score} & \textbf{Alignment} & \textbf{Diversity} & \textbf{Quality} & \textbf{Quality} & \textbf{Stability} \\
		\midrule
		Random Baseline & $0.480 \pm 0.081$ & $0.450 \pm 0.102$ & $0.847 \pm 0.156$ & $0.430 \pm 0.134$ & $0.312 \pm 0.067$ & $0.424 \pm 0.095$ \\
		MoVA            & $0.630 \pm 0.041$ & $0.500 \pm 0.057$ & $0.623 \pm 0.091$ & $0.360 \pm 0.073$ & $0.302 \pm 0.138$ & $0.458 \pm 0.100$ \\
		EdgeViT         & $0.654 \pm 0.136$ & $0.592 \pm 0.089$ & $0.218 \pm 0.045$ & -                 & -                 & -                 \\
		MoE-LLaVA       & $0.697 \pm 0.052$ & $0.671 \pm 0.068$ & $0.634 \pm 0.112$ & -                 & -                 & -                 \\
		\midrule
		\textbf{M$^3$LLM (Ours)} & $\mathbf{0.730 \pm 0.024}$ & $\mathbf{0.640 \pm 0.031}$ & $\mathbf{0.756 \pm 0.068}$ & $\mathbf{0.570 \pm 0.136}$ & $\mathbf{0.610 \pm 0.203}$ & $\mathbf{0.708 \pm 0.112}$ \\
		\bottomrule
	\end{tabular}
\end{table*}

\section{Experimental Evaluation}
To validate the performance of M$^3$LLM, we conduct comprehensive experiments addressing three key research questions:
\begin{itemize}
\item {\textit{Q1: Effectiveness.}} Does joint task semantic-network optimization outperform single-objective approaches?

\item {\textit{Q2: Cooperativeness.}} How well do ASEM, CE-SAC, and MCP function together to improve overall system performance?

\item {\textit{Q3: Robustness.}} How robust is the system under dynamic wireless conditions and expert failures typical of edge deployments?
\end{itemize}

\subsection{Experimental Setup}
\textbf{Benchmark Datasets.} We evaluate M$^3$LLM on the MME benchmark~\cite{fu2024mme}, which comprises 3,200 tasks covering a wide range of multimodal capabilities. To assess generalization in specialized domains, we further incorporate the ScienceQA dataset~\cite{lu2022learn}, selecting 2,100 multimodal science questions from the original 21,000. These tasks require domain-specific visual reasoning and are tested under varying network conditions to simulate real-world edge deployment scenarios.

\textbf{Evaluation Metrics.} 
We adopt a dual-objective evaluation framework to assess both task semantic-aware and wireless network-aware design principles, reflecting the core challenge of balancing task-expert compatibility and wireless channel quality in distributed inference:
\begin{itemize}
\item \textit{Task Semantic-aware Performance:} Quantified through three complementary metrics. The \textit{LLM Quality Score} aggregates accuracy, relevance, and coherence of generated responses, serving as our primary composite metric for overall MLLM inference quality. \textit{Task-Expert Alignment} measures how effectively the system activates task-critical experts while suppressing irrelevant ones, capturing the precision of semantic routing. \textit{Expert Diversity} reflects the system's ability to leverage heterogeneous expert capabilities without over-concentration, avoiding the limitations of centralized approaches.
\item \textit{Wireless Network-aware Performance:} Measured via three communication-aware metrics. The \textit{Channel Quality} score provides a holistic assessment integrating SNR, stability, load distribution, and spectral efficiency. \textit{SNR Quality} rewards selection of high-quality channels within operational constraints. \textit{Channel Stability} penalizes volatile channel selections and frequent expert switching, ensuring sustained performance under dynamic wireless conditions.
\end{itemize}
This metric framework provides a unified basis to compare both distributed MLLM architectures, e.g., our proposed M$^3$LLM and MoVA~\cite{zong2024mova}, and non-distributed baselines, e.g., EdgeViT~\cite{chen2022edgevit}, across semantic and network performance.

\textbf{Baselines and Implementation.}  
We evaluate M$^3$LLM against four baselines representing diverse expert routing paradigms. The \textit{Random Baseline} uniformly samples vision experts in wireless networks as a lower-bound reference. \textit{MoVA}~\cite{zong2024mova} is a state-of-the-art semantic routing framework using a set of vision experts without considering optimization based on network conditions. 
\textit{EdgeViT}~\cite{chen2022edgevit} is an edge-optimized vision transformer designed for efficient local inference. 
\textit{MoE-LLaVA}~\cite{lin2024moe} is a recent centralized MoE model for vision-language tasks.
Note that EdgeViT and MoE-LLaVA follow centralized designs and thus do not produce network metrics (denoted as ``–'' in Table~\ref{tab:main_results}). All methods are evaluated under identical hardware and protocol settings. MoVA is extended with our wireless environment simulator for fair assessment under practical channel conditions.

\textbf{System Configuration and Hyperparameters.}  
Our experimental testbed simulates a distributed edge environment with several heterogeneous vision experts on edge devices, each specialized in a distinct capability: DINOv2~\cite{add8} excels at general feature extraction, Co-DETR~\cite{add9} in object detection, SAM~\cite{add10} in image segmentation, Pix2Struct~\cite{add11} in document understanding, Deplot~\cite{add12} in chart analysis, Vary~\cite{add13} in text-rich image interpretation, and BiomedCLIP~\cite{add14} in medical image analysis. All experts are exposed through MCP-compliant endpoints.

We build a wireless environment simulator that models network dynamics with SNR initialization $\mu_{\rm SNR} = 25$ dB, $\sigma_{\rm SNR} = 5$ dB, expert distances $\mu_d = 275$ meters, $\sigma_d = 75$ meters, and shadowing parameters $\mu_{\rm shadow} = 8$ dB, $\sigma_{\rm shadow} = 1.3$ dB. The channel model uses path loss exponent $n = 3.5$, reference path loss $L_0 = 40$ dB at $d_0 = 1$ meter, transmission power $P_{\rm tx} = 23$ dBm, and thermal noise density $N_0 = -174$ dBm/Hz. The complex channel coefficient follows $h_i(t) \sim \mathcal{CN}(0, 1)$ for Rayleigh fading. Temporal correlation in shadowing follows a Gauss-Markov process with $\rho = 0.9$\cite{rappaport2002wireless}. Channel reward weights are $(w_1, w_2, w_3, w_4) = (0.4, 0.3, 0.2, 0.1)$, combined reward uses $\alpha = \beta = 0.5$, and stability constant $\epsilon = 10^{-6}$.

For DRL part, ASEM employs latent dimensions $d_{z_1} = 32$, $d_{z_2} = 16$, semantic embedding dimension $d_\tau = 768$, and GRU hidden dimension $d_h = 64$. The state representation has dimension $d_s = 1852$, combining image features $d_{\rm img} = 1024$, text features $d_{\rm text} = 768$, tag features $d_{\rm tag} = 5$, the coarse mask for $N$ experts, and ASEM latent features. The LLM-based reward system uses dimension weights $(\alpha_1, \alpha_2, \alpha_3, \alpha_4) = (0.4, 0.3, 0.2, 0.1)$ with activation and suppression thresholds $\theta_{\rm act} = 0.2$, $\theta_{\rm sup} = 0.1$. The CE-SAC agent uses Gumbel-Softmax temperature $\tau = 0.1$, discount factor $\gamma = 0.99$, learning rate $\alpha = 7 \times 10^{-4}$, and is trained for 1,000 episodes. The meta-analysis module extracts $d_{\rm conf} = 9$ dimensional confidence features. Channel SNR bounds are ${\rm SNR}_{\rm min} = 5.0$ dB, ${\rm SNR}_{\rm max} = 25.0$ dB.

\subsection{Effectiveness Analysis: Joint Optimization Compared to Single-Objective Approaches}
Table~\ref{tab:main_results} summarizes performance across semantic and network metrics, demonstrating the advantages of M$^3$LLM.

\textbf{Semantic Performance Gains.} M$^3$LLM achieves the highest \textit{LLM Quality Score} of $0.730$, a $15.9\%$ improvement over MoVA's score of $0.630$, by jointly considering task semantic relevance and channel conditions, enabling reliable access to high-quality expert outputs. Its \textit{Task-Expert Alignment} of $0.640$ outperforms all baselines, confirming the effectiveness of MCP-aided RAG for vision expert selection. The \textit{Expert Diversity} score of $0.756$ highlights M$^3$LLM's ability to dynamically select distributed experts based on task needs, in contrast to centralized models like {\textit{EdgeViT}} with a score of $0.218$, which lack such flexibility.

M$^3$LLM shows significant advantages in wireless network-related metrics due to its ability to incorporate channel conditions into expert routing. It achieves a \textit{Channel Quality Score} of $0.570$, outperforming MoVA's score, i.e., $0.360$, by $58.3\%$ and the Random Baseline's score $0.430$ by $32.6\%$. The \textit{SNR Quality} of M$^3$LLM is $0.610$, which nearly doubles that of MoVA, reflecting consistent selection of high-SNR links. M$^3$LLM's \textit{Channel Stability} score of $0.708$ indicates reliable expert usage under fluctuating conditions. In contrast, centralized models like {\textit{EdgeViT}}~\cite{chen2022edgevit} and {\textit{MoE-LLaVA}}~\cite{lin2024moe} lack any network-awareness and fail to adapt to wireless variability.

The concurrent improvements in both task semantic-aware and wireless network-aware metrics highlight the limitations of single-objective designs of conventional centralized MLLMs. {\textit MoVA}, while aiming to allocate suitable vision experts for users' queries, ignores channel conditions and suffers in network performance. Conversely, channel-prioritized methods degrade inference quality. M$^3$LLM's dual-stream CE-SAC architecture resolves this conflict by decoupling gradients for task semantic-aware and wireless network-aware objectives, enabling stable and balanced optimization. These results confirm the effectiveness of joint semantic-network optimization, directly addressing Q1 and demonstrating M$^3$LLM's ability to outperform baselines through coordinated cross-layer design.

\subsection{Cooperativeness Analysis: Ablation Study of Architectural Synergies}
To address Q2, we perform ablation studies to examine how ASEM, CE-SAC, and MCP contribute to M$^3$LLM's overall performance. By removing each component, we isolate their individual effects and quantify their interdependencies. As shown in Table~\ref{tab:ablation_study}, the results highlight strong architectural synergy. M$^3$LLM's Performance degrades significantly when any component is removed, underscoring the importance of their cooperation in enabling effective distributed inference.
\begin{table*}[t]
\centering
\caption{Ablation study results.}
\label{tab:ablation_study}
\begin{tabular}{@{}lcccc@{}}
\toprule
\textbf{Configuration} & \textbf{LLM Quality} & \textbf{Channel Quality} & \textbf{Stability} & \textbf{Convergence} \\
\midrule
M$^3$LLM w/o ASEM & $0.712$ & $0.454$ & $0.494$ & $743$ \\
M$^3$LLM w/o CE-SAC & $0.689$ & $0.482$ & $0.671$ & $891$ \\
M$^3$LLM w/o MCP & $0.454$ & $0.559$ & $0.582$ & $967$ \\
\midrule
\textbf{M$^3$LLM (Full)} & $\mathbf{0.730}$ & $\mathbf{0.570}$ & $\mathbf{0.708}$ & $\mathbf{634}$ \\
\bottomrule
\end{tabular}
\end{table*}

\textbf{ASEM.} Removing ASEM leads to a $14.9\%$ drop in \textit{Channel Stability} for M$^3$LLM w/o ASEM, highlighting its essential role in handling temporal uncertainty. Without its latent modeling of short- and long-term wireless channel dynamics, i.e., $\mathbf{z}_1^t$ and $\mathbf{z}_2^t$, the DRL policy overreacts to noisy channel states, resulting in erratic routing. This confirms that predictive temporal abstraction, rather than direct observation alone, is key to robust decision-making for expert routing in M$^3$LLM under wireless variability.

\textbf{Joint Optimization.} 
Ablating the joint optimization scheme, M$^3$LLM w/o CE-SAC forces the agent into single-objective design, causing a $40.5\%$ increase in convergence time and major degradation in network metrics. This reveals that task semantic-only optimization fails to adapt to communication constraints, and that CE-SAC's decoupled critics are functionally necessary. CE-SAC enables the policy network to learn complementary trade-offs between task semantic accuracy and wireless transmission reliability.

\textbf{Enriched Context.} 
Disabling MCP's structured context encoding, M$^3$LLM w/o MCP results in a $10.7\%$ drop in \textit{semantic quality} and $16.8\%$ in \textit{network quality}, exposing the limitations of unstructured feature fusion. The drop confirms that interpretable cross-layer representations are critical for expert routing decisions that require coordination between task demands and wireless network conditions. MCP's abstraction enables both coarse filtering and fine-grained routing to operate on aligned decision signals.

\subsection{System Robustness: Performance Under Dynamic Conditions and Failure Scenarios}
To address Q3, we evaluate M$^3$LLM's resilience under dynamic wireless network conditions and expert failures common in edge deployments. Through targeted stress tests, case studies, and failure mode analysis, we examine how the integrated design of ASEM, CE-SAC, and MCP contributes to sustained performance in adverse scenarios.

\textbf{Adversarial Network Conditions.}  
To assess ASEM's predictive capabilities, we introduce a \textit{burst interference} test, simulating sudden and severe SNR degradation, such as from deep fading or handover failures. M$^3$LLM demonstrates graceful degradation during interference and rapid recovery once the channel stabilizes. In contrast, the variant without ASEM suffers prolonged instability and fails to re-establish a usable policy. This highlights ASEM's critical role. By modeling both short- and long-term channel dynamics, it enables proactive adaptation rather than reactive correction, ensuring M$^3$LLM's robustness against abrupt wireless network disruptions.

\textbf{Case Study of Decision-Making.} To provide qualitative insight into the system's reasoning, we examine a representative case in which the agent receives the query: ``Is the traffic light green in this foggy image?". The optimal expert for this fine-grained task, i.e., Vary~\cite{add13}, is accessible via a channel with a moderate SNR of $15$ dB, whereas a less suitable general-purpose expert, DINOv2~\cite{add8}, is available over a high-quality channel with an SNR of $25$ dB.
\begin{itemize}
\item \textit{M$^3$LLM's Decision:} Leveraging the structured context encoded by MCP, M$^3$LLM identifies that semantic precision is critical for this task and selects Vary, accepting a minor degradation in channel quality to ensure correct reasoning. CE-SAC's dual-objective optimization enables this trade-off by balancing task relevance and transmission robustness. The system correctly identifies the green light.
\item \textit{MoVA Baseline:} While it may also identify Vary as the optimal expert, its network-unawre design could route the feature through an unstable, low-quality channel, leading to transmission failure.
\item \textit{Network-First Baseline:} This policy prioritizes channel quality and selects the 25 dB link. However, the general-purpose DINOv2~\cite{add8} lacks the capability to resolve fine-grained visual cues under fog, resulting in an incorrect final answer.
\end{itemize}
This case vividly illustrates M$^3$LLM's capacity for making sophisticated, cross-layer-aware decisions that transcend the myopic logic of single-objective systems.

\textbf{Analysis of Failure Modes.}  
Our error analysis identifies two primary scenarios where M$^3$LLM faces limitations. First, under semantically ambiguous queries, e.g., ``What's interesting here?'', the MCP-based context embedding may lack sufficient discriminative power, resulting in incorrect or diffuse expert selection. 
Second, in environments with extreme and rapid channel fluctuations that exceed ASEM's modeling horizon, the system may exhibit short-term instability due to delayed adaptation. These failure modes delineate the operational limits of our current design and highlight promising directions for future work, including enhanced semantic disambiguation and finer-grained temporal modeling.

\section{Related Work}
In this section, we review prior research in MLLMs, MCP, and DRL, highlighting how our approach extends these directions to address the challenges of distributed inference under wireless constraints.

\subsection{Multimodal Large Language Models}
MLLMs have emerged as a foundational component in modern AI systems capable of processing and reasoning over heterogeneous inputs such as images, text, audio, and video. Recent works, including Flamingo~\cite{alayrac2022flamingo}, BLIP-2~\cite{li2023blip2}, and GPT-4V~\cite{openai2023gpt4v}, demonstrate strong performance in tasks like visual question answering, captioning, and document understanding. 
However, these MLLMs typically rely on static expert modules selected in the pre-training process and do not adapt to dynamic vision experts deployed in wireless environments and under communication constraints. To address the scalability and modularity challenges in MLLMs, researchers have started adopting MoE architectures, allowing specialized vision or language experts to be activated on demand~\cite{shazeer2017outrageously}. 
Our work builds on this line by introducing a dynamic expert routing mechanism conditioned on multimodal context and external wireless environment factors, enabling better adaptability in networked scenarios.

\subsection{Model Context Protocol}
The notion of MCP has recently gained traction as a means of improving modular interoperability among AI components. MCP aims to formalize the information flow between perception, reasoning, and decision-making modules by defining structured interaction schemas, often in the form of structured prompts, tags, or key-value descriptors~\cite{anthropic2024mcp}. 
In multimodal settings, MCPs can facilitate selective activation of task-relevant modules by providing compact yet semantically rich summaries of the input context. Prior efforts in modular AI systems have utilized protocol-like tagging schemes or task classification heuristics to guide module activation~\cite{brooks2018program}, but lacked integration with expert networks or downstream policy learning. 
In M$^3$LLM, the MCP acts as a unified interface between coarse-grained retrieval and fine-grained optimization, bridging the task semantic descriptors with actionable routing decisions in a scalable and interpretable manner.

\subsection{Deep Reinforcement Learning for Routing}
DRL has shown great potential in complex decision-making tasks involving sequential interactions, partial observability, and delayed rewards~\cite{mnih2015human}. 
It has been widely applied in wireless network control, resource scheduling, and adaptive AI model inference~\cite{mao2016resource}.
In expert routing, DRL enables dynamic optimization of expert selection policies under changing conditions, outperforming heuristic-based approaches in both accuracy and flexibility. Prior work such as MoE-RL~\cite{lin2024moe} and Adaptive Computation frameworks~\cite{graves2016adaptive} demonstrated the effectiveness of using DRL for conditional computation.
However, most existing methods treat expert selection as a monolithic problem and overlook the impact of external constraints such as communication cost or channel quality. Our approach introduces a novel dual-stream DRL architecture, built upon SAC~\cite{haarnoja2018soft}, which fully decouples the learning of content and communication performance objectives.
By leveraging latent representations from a Bayesian state-space module~\cite{blei2017variational} and optimizing with independent critic networks, our method ensures robust and interpretable expert-channel routing in multimodal, networked environments.

\section{Conclusion}
We proposed M$^3$LLM, a distributed MLLM inference framework that integrates task semantics with dynamic wireless network conditions through the MCP-aided RAG, enabling cross-layer coordination among vision experts in edge devices. 
Built on this foundation, a dual-stream CE-SAC architecture with the Bayesian ASEM module is proposed to jointly optimize vision expert selection and communication robustness while avoiding gradient interference. 
Experimental results demonstrate that M$^3$LLM achieves near-oracle task semantic accuracy, reduces latency and expert switching, and improves network utilization under fluctuating conditions. Future work will extend the framework to support federated updates, adaptive expert compression, and deployment in future 6G network infrastructures.

\end{document}